\documentclass[twoside]{article}
\usepackage{fleqn,espcrc2}
\usepackage{epsf}
\newcommand  {\rf} [1]{(\ref{#1})}
\newcommand  {\beq}[1]{\begin{equation}\label{#1}}
\newcommand  {\eeq}   {\end{equation}}
\newcommand  {\bea}   {\begin{eqnarray}}
\newcommand  {\eea}   {\end{eqnarray}}
\newcommand  {\e}     {\mbox{e}}
\renewcommand{\d}     {\mbox{d}}
\renewcommand{\d}     {\delta}
\newcommand  {\g}     {\gamma}

\renewcommand{\l}     {\lambda}

\renewcommand{\t}     {\tau}
\renewcommand{\b}     {\beta}

\newcommand  {\n}     {\nu}

\newcommand  {\R}     {{\rm I\!R}}

\newcommand\npb  [3]{{{\it Nucl.\ Phys.\ }{\bf B #1} (#2) #3}}

\newcommand{\hepth}  [1]{{\tt hep-th/#1}}

\title{On the phase diagram of 2d Lorentzian Quantum Gravity}
\author{Jan Ambj\o rn$^{\rm a}$, 
{\bf K. N. Anagnostopoulos}\address{
Niels Bohr Institute, Blegdansvej 17, DK--2100 Copenhagen \O, Denmark}
and
R. Loll\address{Albert-Einstein-Institut,
Am M\"{u}hlenberg 5, D-14476 Golm, Germany}}

\begin{document}

\begin{abstract}
The phase diagram of 2d Lorentzian quantum gravity (LQG) coupled to
conformal matter is studied. A phase transition is observed at
$c=c_{\rm crit}$ ($1/2<c_{\rm crit}<4$) which can be thought of as the
analogue of the $c=1$ barrier of Euclidean quantum gravity (EQG). The
non--trivial properties of the quantum geometry are discussed.

\vspace{1pc}
\end{abstract}
\maketitle
\section{INTRODUCTION}
Recently a new model of 2d quantum gravity has been proposed
\cite{al}. It is defined using dynamical triangulations from a subclass
of diagrams which can be given a causal structure. Such diagrams are
generated by gluing together one dimensional time--slices or
``universes'' (in our case a set of vertices connected by space-like
links forming a diagram with the topology of a circle) with time-like
links such that they form a triangulated surface.  Vertices
connected by time-like links are causally related and a unique time can
be assigned to the vertices of each time--slice.  Such graphs can be
given a Lorentzian metric by defining time-like links to have equal
negative length squared and space-like to have positive. All triangles
have equal area and the volume of spacetime is proportional to the
number of triangles $N_T$. The system has been found to have a
non--trivial continuum limit only at an imaginary value of the
cosmological constant $\l$. The geometry of space is maximally
fluctuating but the system is much smoother than Liouville gravity: By
defining the two point function to be
\beq{tpt}
G(\l,t)=\sum_{T\in{\cal T}} \e^{-\l N_T}\, ,\quad \l\in\R \, ,
\eeq 
where the summation is over triangulations $T$ of cylindrical topology
with $t$ time--slices, one finds that the Hausdorff dimension $d_H$
of the system is $2$. This is entirely due to the imposition of the
causal structure: If one allows the creation of baby universes the
system becomes the ordinary EQG model.  If conformal
matter of central charge $c$ is coupled to the system, it has been
found in the
case of one Ising model \cite{aal} that the coupling is weak for $c\le
1/2$ and the bulk properties of geometry do not change
(e.g. $d_H=2$). The
critical matter system belongs to the Onsager universality class.
When 8 Ising spins are coupled to gravity, a qualitatively different
behaviour emerges. The system undergoes a phase transition for some
$1/2<c_{\rm crit}<4$ \cite{AAL} to a phase where we observe
anomalous scaling between the typical length scale $L$ and typical
time scale $T$. It is found that ${\rm dim}[L]=2\, {\rm dim}[T]$ and
that $d_H=3$. The latter is a ``cosmological'' Hausdorff
dimension. The former is a relation which has been found to hold 
for the non--singular part of the quantum geometry 
of EQG for {\it any} value of $c\le 1$. The short
distance behaviour of space--time is given by a different fractal
dimension $d_h=2$. The quantum geometry is
different at different scales and has a more complex structure. 
The matter critical exponents are Onsager.
We conclude that the matter coupling to geometry
in our model is much weaker than in the case of EQG so that the large
$c$ phase has 
non--degenerate, interesting continuum limit for the quantum
geometry. Nothing dramatic happens from the point of view of matter
but the geometry undergoes a qualitative change and even shares some
features with the non--singular part of EQG in the $c<1$
phase. This is the first time where one has a model which
demonstrates explicitly that the strong coupling of matter and
geometry in Liouville gravity is {\it entirely} due to the ({\it a
priori}) presence of baby--universes.

\section{NUMERICAL RESULTS}
\begin{figure}[t]
\centerline{\epsfxsize=8.0cm\epsfysize=5.333cm\epsfbox{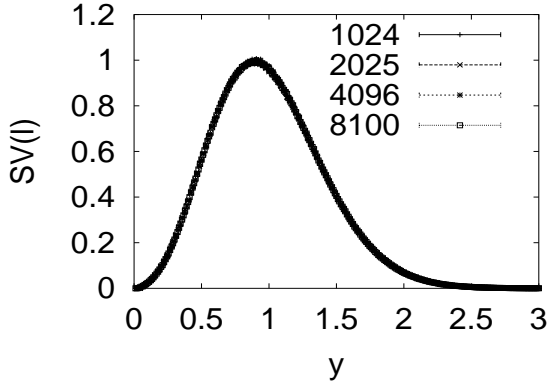}}
\caption{The $SV(l)$ distributions for the $c=1/2$, $\t=1$ system. 
$y=l/N^{1/\d_h}$}
\label{f:1}
\end{figure}

\begin{figure}[t]
\centerline{\epsfxsize=8.0cm\epsfysize=5.333cm\epsfbox{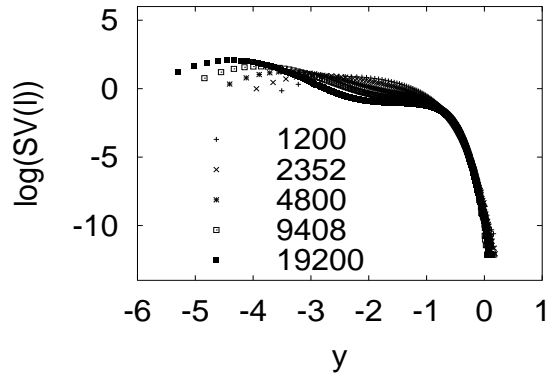}}
\caption{The $SV(l)$ distributions for the $c=4$, $\t=3$ system.
$y=l/N^{1/\d_h}$}
\label{f:2}
\end{figure}






We simulate triangulated surfaces with the topology of the torus with
$N_T$ triangles and $N=N_T/2$ vertices \cite{aal,AAL}. The temporal
length (number of time slices) is $t$ and we choose $N=t^2/\t$ for
$\t= 1,2,3$ and $4$. The geometry is updated with the move described
in \cite{aal} and the Ising spins with the Swendsen--Wang
algorithm. The partition function for $n$ Ising models is
\beq{2:1}
G(\l,t,\b)=\sum_{T\in {\cal T}} {\rm e}^{-\l N_T} Z^n_T(\b)\, .
\eeq
$Z_T(\b)$ is the partition function of an Ising model with spins
placed at the vertices of the graph $T$ at inverse temperature
$\b$. $n=1,8$ corresponding to $c=1/2,4$. We obtain the fixed volume
partition function by adding a gaussian volume--fixing term and
measuring only on configurations of given volume $N_T$. The critical
values of $(\l,\b)$ are determined. For $c=0$ they are $\l_c=\ln 2$,
for $c=1/2$ they are $(\l_c,\b_c)= (0.742(5),0.2521(1))$ and for $c=4$
they are
$(\l_c,\b_c)= (1.081(5),0.2480(4))$ ($\l\rightarrow \ln 2$ when
$\b\rightarrow\infty$). Those values are insensitive to $\t$. Finite
size scaling (FSS) is applied at $(\l_c,\b_c)$ in order to measure the
system's scaling properties.

\begin{figure}[t]
\centerline{\epsfxsize=8.0cm\epsfysize=5.333cm\epsfbox{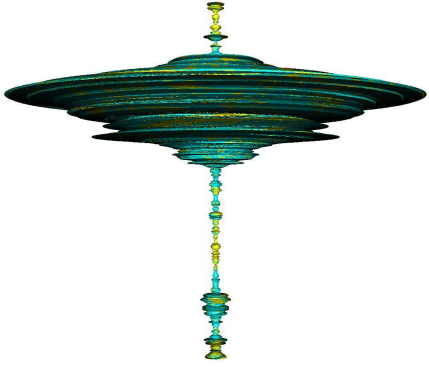}}
\caption{A typical configuration of the $c=1/2$, $\t=3$ system. $N_T=73926$.}
\label{f:3}
\end{figure}

The first quantity that we measure is the distribution $SV(l)$ of
spatial volumes $l$. We expect a scaling behaviour
\beq{2:2}
SV_N(l) = F_S(l/N_T^{1/\d_h})\, ,
\eeq
for some function $F_S$. $\d_h$ is related to the fractal dimension
$t\sim N_T^{1/d_H}$ by $d_H=\d_h/(\d_h-1)$. Here, $t$ is the
dynamically generated time extend of the scaling part of
space--time. For the $c=1/2$ model we find $\d_h=d_H=2$ as shown in
Fig.~\ref{f:1}. For $c=4$ the scaling behaviour of $SV(l)$ is
qualitatively different. As can be seen in Fig.~\ref{f:3}, 
the configurations show a tendency to form a
long and thin neck with spatial volume of the order of the cutoff and
an extended region that scales according to \rf{2:1}. The volume of the
extended part is asymptotically proportional to $N_T$. The effect is
seen for large enough volumes and finite size effects are minimised by
taking $\t\ge 3$. For $\t=3$ we find $\d_h=1.54(3)$ (see
Fig.~\ref{f:2}) and for $\t=4$
$\d_h=1.50(3)$. Let us assume for clarity that $\d_h=3/2$. Then
$d_H=3$ and $l\sim t^2$ where $l$ is the typical length scale of
universes in the extended region and $t$ the typical time scale of their
existence. 

\begin{figure}[t]
\centerline{\epsfxsize=8.0cm\epsfysize=5.333cm\epsfbox{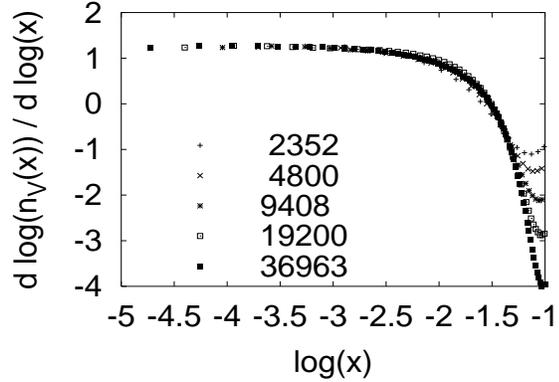}}
\caption{Short distance scaling of $n_N(r)$ for $c=4$, $\t=3$.}
\label{f:4}
\end{figure}
\begin{table}[b]
\caption{The critical exponents for the $c=1/2$, $\t=1$ and the $c=4$,
$\t=3$ systems.}
\label{t:1} 
\renewcommand{\tabcolsep}{1pc} 
\renewcommand{\arraystretch}{1.2} 
\begin{center}
\begin{tabular}{ c  c c c }
\hline
                &                   & $c=1/2$ & $c=4$   \\
\hline
 $\chi$         &$\frac{\g}{\n d_H}$& 0.89(1) & 0.85(1) \\
 $D_{\ln |m|}$  &$\frac{1} {\n d_H}$& 0.526(5)& 0.520(5)\\
 $D_{\ln m^2}$  &$\frac{1} {\n d_H}$& 0.525(5)& 0.512(5)\\
\hline
\end{tabular}
\end{center}
\end{table}

The above results are supported by measuring the volumes $n_N(r)$ of
geodesic spherical shells at distance $r$. We expect that 
$n_N(r)=N_T^{1-1/d_H}F_1(x)$, $x=r/N_T^{1/d_H}$ which is known to hold
in EQG for {\it all} $c$. $d_H$ is the {\it cosmological}
Hausdorff dimension describing how volume and time have to scale in
order to obtain a non trivial continuum limit of the two loop
function. On the other hand the short distance behaviour is given by
$d_h$, where $n_N(r)\sim r^{d_h-1}$ at scales $r\ll N^{1/d_H}$. For
$c=0$ we find analytically that $d_H=d_h=2$ and for $c=1/2$
numerically that $d_H=d_h=2.00(5)$. For $c=4$, $n_N(r)$ has different
scaling behaviour at different length scales. For $r\ll N^{1/3}$ we
find that $d_h=2.1(2)$ as can be seen of Fig.~\ref{f:4}. For $r\gg
N^{1/3}$ the value of the tail of $n_N(r)$ is almost independent of
$N_T$ and $r$ showing dominance of 1d configurations. For $r\sim
N^{1/3}$ the scaling of the peaks of $n_N(r)$ gives $d_H=3.07(9)$. 

The matter scaling exponents are computed from the scaling behaviour
of the magnetic susceptibility $\chi\sim N^{\gamma/\nu d_H}$ and
$D_{\ln |m|} \equiv \frac{d \ln |m|}{d \b}\sim N^{1/\n d_H}$ and
$D_{\ln m^2} \equiv \frac{d \ln m^2}{d \b}\sim N^{1/\n d_H}$. The
$c=1/2$ system clearly belongs to the Ising universality class. For
$c=4$ special care has to be taken in order to isolate the critical
behaviour of the spin system in the extended region of space--time
\cite{AAL}. The critical exponents are consistent with Onsager
values. Our results are summarized in Table~\ref{t:1}.




\end{document}